\documentclass[twocolumn,showpacs,aps]{revtex4}
\usepackage[dvips]{graphicx}
\usepackage{bm}

\begin{document}

\title{\textbf{ Renormalized quantum Fisher information manifestation of
\\
Berezinskii-Kosterlitz-Thouless phase transition for spin-$\frac{1}{2}$ XXZ chain }}


\author{Qiang Zheng$^{1,~2}$, Yao Yao$^{1}$, Xun-Wei Xu$^{1}$, Yong Li$^{1}$}
\address{
$^{1}$ Beijing Computational Science Research Center, Beijing 100084, China
\\
$^{2}$ School of Mathematics and Computer Science, Guizhou Normal
University, Guiyang, 550001, China
}

\begin{abstract}
Combining the ideas of quantum Fisher information and quantum renormalization group method,
the Berezinskii-Kosterlitz-Thouless quantum phase transition of spin-$\frac{1}{2}$ XXZ chain is investigated.
Quantum Fisher informations of the whole $N$ sites and the partial $\frac{N}{3}$ sites are studied. They
display very similar behaviors, even though their mathematical formulas are very different from each other.
The universally critical exponent of quantum Fisher information is obtained as $\beta=0.47$, which is
consistent with the results obtained by the renormalized concurrence or discord. We also discuss the relationship between quantum Fisher information and entanglement. 
\end{abstract}


\maketitle

\section{Introduction}
Quantum phase transition (QPT) \cite{Sachdev} takes place at zero
temperature. For QPT, only quantum fluctuation plays a
role and the thermal fluctuation vanishes. At a critical point,
the property of ground state changes qualitatively.
Nonanalytic behavior of a physical quantity and its corresponding scaling laws
can be used to characterize the second-order QPT.
The set of critical exponents \cite{Barber} associated
with the scaling laws represents the important information of
transition, which are also used to classify the majority of critical systems.

The spin chain plays an important role in the study of QPT \cite{txiang08}.
Recently, in addition to order parameter and symmetry broken, which are
two key concepts in Landau-Ginzburg-Wilson theory, quantum
entanglement and its scaling laws of spin chain have been proposed to detect the
critical point of QPT \cite{Osterloh, Osborne, LAWu, vidal, Hofmann14}.
Near the critical point, a long-range correlation develops.
Note most operable measures of entanglement such as concurrence
can only describe the short-range correlation. In order to better investigate
the long-range correlation property of QPT, combining the ideas of
quantum renormalization group \cite{kgwil} and
concurrence, the notion of \textit{renormalization of concurrence} was introduced in
Ref.~\cite{Kargarian07, Kargarian08b, Kargarian09, fwma11}.
Recently, quantum discord \cite{Sarandy09} and Clauser-Horne-Shimony-Holt
inequality \cite{batle10, Justino12, jlchen12} are also adopted
to predict QPT of spin chain.

All the upper mentioned physical quantities, such as concurrence or discord,
describe quantum correlation of system.
Additionally, some strongly correlated quantum many-body systems, such as quantum Hall system, has
a topological order \cite{xgwen04}, which depends on the topology of system
and do not have local-order parameters. These facts motivate us to ask:
without using these local quantum correlations, it is possible to probe 
the Berezinskii-Kosterlitz-Thouless(BKT) QPT \cite{kt73}. For BKT QPT, the
correlation length is divergent, but the correlation function decays algebraically
with distance, in contrast to usually QPT with a exponential decay of correlation.
The BKT transition is a topological phase transition with vortex-antivortex pairs
dissociating above the critical point in a two-dimensional system \cite{aalt06}.
There is still BKT QPT for one-dimensional spin-$\frac{1}{2}$ chain \cite{Giamarchi03}.

On the two sides of critical point of QPT, the property of the
ground states changes qualitatively. Note also that
QFI describes how well small change of a parameter can be probed. Therefore,
we will show that QFI with respect to the external parameter inducing the QPT
can be used to probe BKT QPT. 
QFI, obtained by extending Fisher information \cite{Fisher}
from classical regime to quantum regime,
is introduced in quantum estimation theory \cite{Helstrom}.
It is related to Cram\'{e}r-Rao inequality \cite{Holevo},
which determines the low bound of the optimal quantum estimation.
Recently, QFI has inspired wide interests due to its own
importance in the quantum estimation and
quantum metrology \cite{Escher11, Lloyd11, Rivas, Boixo08, Ferrie13, sanders11, jin13}. It
can be used to characterize non-Markovianity of
environment \cite{Wang10b}, Heisenberg uncertainty relationship \cite{luo03b, Ueda11},
and entanglement \cite{Smerzi09, Luo13b}.

Actually, QFI has been used to 
characterize the QPT of Lipkin-Meskhov-Glick model \cite{jma09},
spin-$\frac{1}{2}$ XY model\cite{WFLiu13} and
a quantum-critical spin chain environment \cite{zsun10}, and the dynamical phase transition
of Bose-Einstein condensate in the double well potential \cite{wzhong12a}.
Recently, it has been reported that the QPT of Dicke model has been sensitively probed by QFI \cite{Jin13a}.
The QPT can be considered as a useful resource to enhance the parameter estimation \cite{Invernizzi08}
as the value of QFI near the critical point of QPT is larger than one outside the critical regime.

In this paper, combining the idea of QFI and quantum renormalization group method \cite{kgwil},
we investigate the BKT  QPT 
of one-dimensional spin-$\frac{1}{2}$ XXZ chain
at the isotropic antiferromagnetic point $\Delta=1$. 
We find that in the two phases of XXZ model,
QFI shows different behaviors.
In the vicinity of the critical point, the first-order derivative of the QFI with respect to the anisotropy parameter is singular. The corresponding finite-size scaling laws are obtained numerically,
with the universal critical exponent $\beta=0.47$. It equals to the results gained
by the renormalized concurrence \cite{Kargarian08a} or discord \cite{zfh12}.
Compared to the method of the renormalized \cite{Kargarian08a} or discord \cite{zfh12},
the advantage of the renormalized QFI is that it can be used beyond the effectively
two-qubit case.

\section{Motivation}
This section discusses the motivation of our paper.
The Bures distance is defined as \cite{Nielsen10}
$\mathcal{D(\rho, \sigma)}^2=2[1- \mathcal{F(\rho, \sigma)}]$,
with $ \mathcal{F(\rho, \sigma)}= \mathrm{Tr}(\sqrt{\rho} \sigma \sqrt{\rho}) $ being  the fidelity.
The Bures distance or fidelity describes the closeness
of two quantum state $\rho$ and $\sigma$ in Herbert space.
It is also related to QFI $F_{\varphi}$ (defined in latter Eq.~(\ref{deffisher}))
by \cite{caves94}
\begin{equation}
\begin{array}{llll}
\mathcal{D}(\rho_{\varphi}, \rho_{\varphi+ d \varphi})^{2}=\frac{1}{4}F_{\varphi} d \varphi^{2}.
\end{array}
\end{equation}
Here $\varphi$ is a parameter to be estimated.
This relationship shows that QFI quantifies how well small change of a parameter can be probed.

It is well-known that for a QPT of a quantum many-body system,
the ground states changes drastically on the two sides of critical point.
Therefore, two kinds of fidelity, the ground states overlap \cite{Zanardi06}
and Loschmidt echo \cite{htquan06}, are used to characterize the QPT.
The main advantage of this approach is that the
fidelity is just a purely Hilbert-space geometrical quantity. Moreover,
the fidelity susceptibility (FS) \cite{You07}
$\chi_{F}=-\frac{\partial^{2} \mathcal{D}(\rho_{\varphi}, \rho_{\varphi+ d \varphi})}{\partial \varphi^{2}}$,
can also manifest the quantum critical point \cite{You08}. Very recently, it is proved rigorously that \cite{xgwang14a} for any state the QFI is proportional to the fidelity susceptibility.
For example, the QFI of a pure state $|\psi \rangle$ is
$F_{\varphi}=4[\langle \partial_{\varphi} \psi | \partial_{\varphi} \psi \rangle - |\langle \psi | \partial_{\varphi} \psi \rangle|^{2} ]$, which is proportional to the FS of
$|\psi \rangle$ with respect to the parameter $\varphi$.

However, there is discrepancy on whether FS
could be used to probe the BKT transition of XXZ model at $\Delta=1$.
Using Luttinger-liquid description, Yang \cite{yang} and Fjaerestad \cite{fjaer} had obtained the analytical expression of the ground state FS, which is divergence at the BKT transition point. However,
the finite-size scaling can not been obtained by this method \cite{yang, fjaer}.
Later, Wang \textit{et.al.} \cite{mfeng10} using the density-matrix-renormalization-group technique numerically
had uncovered this critical point.
In contrast to these papers,
Chen \textit{et. al.} \cite{chens08} had \textit{not} found the FS singularity at the
BKT transition point with exact diagonaliztion method for the relative small system.

These facts inspire us to revisit this issue and adopt the QFI to manifest the quantum critical point.
The ground-state overlap fidelity combined with the quantum RG method \cite{Rezakhani12B}
has been adopted to investigate the QPT of spin chain.
However, it is still an open question that whether
the renormalized QFI (or FS) can detect the critical point of BKT QPT.
This is the aim of our paper.

\section{Renormalized spin-$\frac{1}{2}$ XXZ model}
The renormalized group (RG) \cite{kgwil}, with key idea to
reduce degree of freedom by a recursive procedure,
is a crucial method to study the QPT.
Originally, Kadanoff divided the lattices of one dimensional spin chain into blocks,
and treated every block as a new lattice. The lowest eigenvectors of the block are
used to construct the projectional operator. With this projectional
operator, the original Hamiltonian is projected
onto the renormalized space step by step.
As a result, an effective Hamiltonian is obtained,
which has the same mathematical form to the original one and
the parameters are renormalized.

As an example, we study a spin-$\frac{1}{2}$ XXZ model on a periodic chain of $N$ sites.
Its Hamiltonian reads to be \cite{Kargarian08a}
\begin{equation}
\begin{array}{llll}
H=\frac{1}{4}J \sum_{i}^{N} (\sigma_{i}^{x}\sigma_{i+1}^{x}+
\sigma_{i}^{y}\sigma_{i+1}^{y}+ \Delta
\sigma_{i}^{z}\sigma_{i+1}^{z}),
\end{array} \label{ham1}
\end{equation}
where $J >0$ is the exchange coupling parameter, $\Delta >0$ is the anisotropy
parameter, and $\sigma_{i}^{\alpha}$ $(\alpha=x,y,
z)$ are the Pauli matrices at site $i$.
The energy spectrum of 1D XXZ model can be
exactly solved by by Bethe ansatz \cite{cnyang}.
For $\Delta=1$, the Hamiltonian has SU(2) symmetry, while for $\Delta\neq
1$, SU(2) symmetry breaks down to U(1) rotation symmetry around the $z$ axis.
In particular, the correlation function at $\Delta=1$ behaves as
\cite{rorbach58}
$|\langle \sigma_{i}^{z}\sigma_{i+R}^{z} \rangle| \propto \frac{1}{R}$ as
$R \rightarrow \infty$, showing there is quasi-long range order.
Therefore, $\Delta= 1$ is the critical point of BKT transition.

To show how the quantum RG method works,
a block composed by a three-site is chosen. Choosing an odd sites block is critical for XXZ model
because this makes each block with two degenerate ground states.
If the two ground states are adopted to constitute the
projectional operator,
after RG the Hamiltonian preserves a self-similar form as the original one.
Following Ref.~\cite{Kargarian08a}, the original Hamiltonian is divided into two parts
\begin{equation}
H=H_{B}+H_{BB}.
\end{equation}
Here the intrablock three-sites Hamiltonian is
\begin{equation}
\begin{array}{llll}
H_{B}=\sum_{I}h_{I}^{B}
\\
=\sum_{I} \frac{1}{4}J (\sum_{k=1,
2} (\sigma_{I, k}^{x}\sigma_{I, k+1}^{x}+ \sigma_{I,
k}^{y}\sigma_{I, k+1}^{y}+ \Delta \sigma_{I, k}^{z}\sigma_{I,
k+1}^{z}),
\end{array}
\end{equation}
and the interblock coupling one is
\begin{equation}
\begin{array}{llll}
H_{BB}=\sum_{I}h_{I}^{BB}
\\
=\sum_{I} \frac{1}{4}J (\sigma_{I,3}^{x}\sigma_{I+1, 1}^{x}+
\sigma_{I,3}^{y}\sigma_{I+1, 1}^{y}+ \Delta
\sigma_{I,3}^{z}\sigma_{I+1, 1}^{z})
\end{array}
\end{equation}
with $I$ being index of block.

\begin{figure}[!tb]
\centering
\includegraphics[width=1.5in]{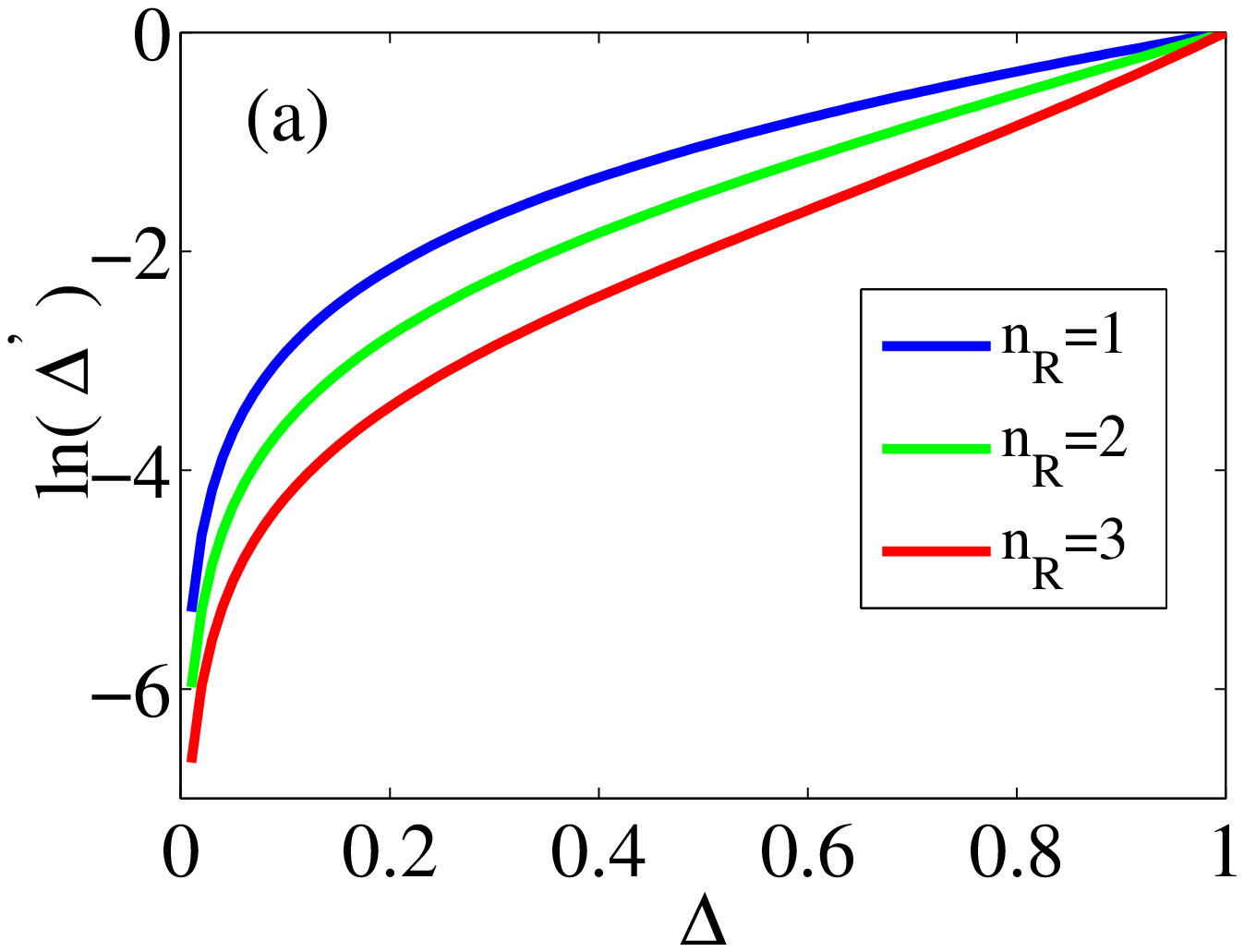}
\includegraphics[width=1.5in]{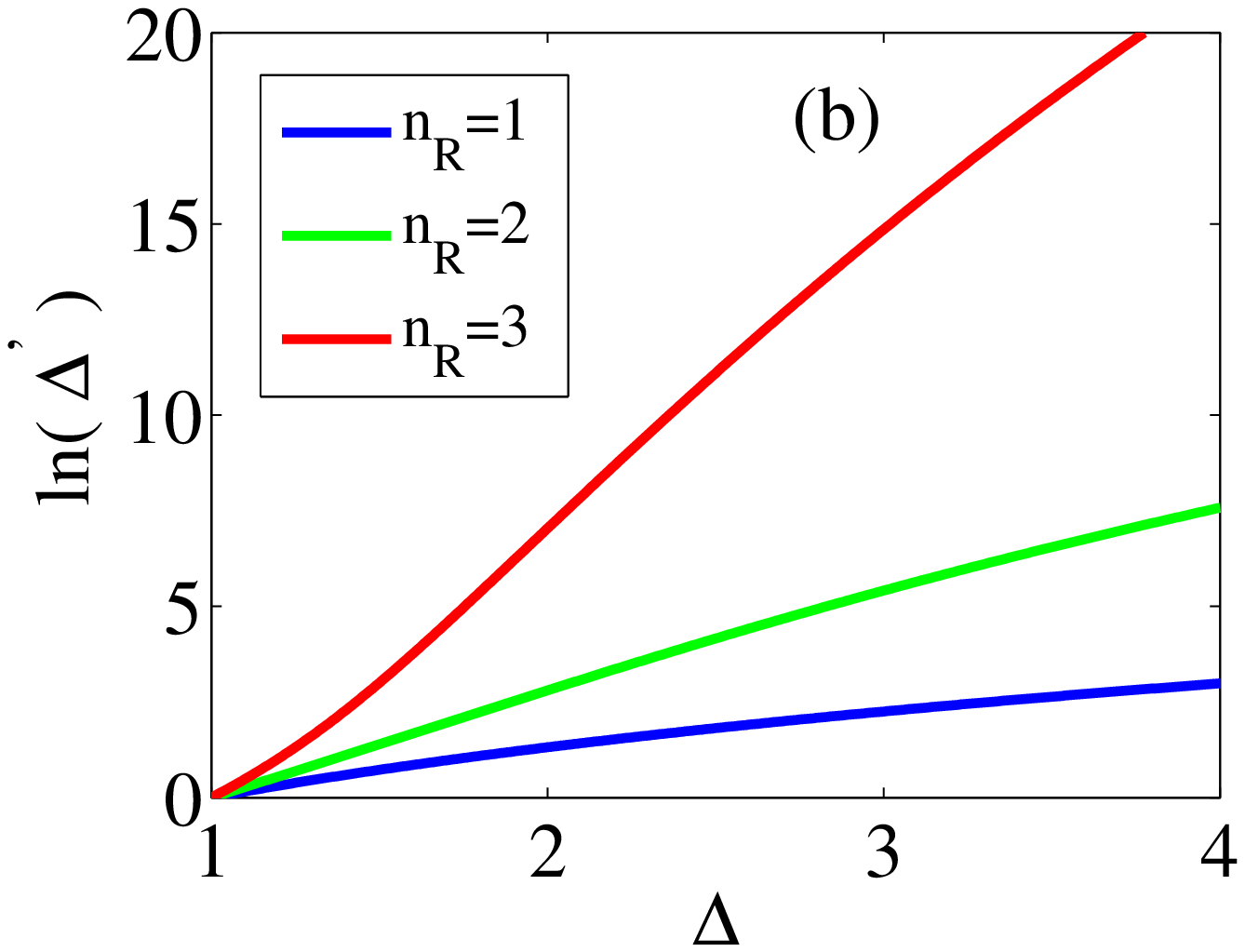}
\hspace{2.2cm}\caption{
The variation of $\ln(\Delta'$) with respect to
$\Delta$. Here $\Delta'$ is the renormalized anisotropy parameter defined in Eq.~(\ref{iterrel}).
(a) and (b) corresponds to the spin-liquid phase and N\'{e}el phase, respectively.}
\label{evoldelta}
\end{figure}

The eigenvectors and eigenvalues of each three-site block $h_{I}^{B}$ can be obtained by
exact diagonalization. In order to construct the
projectional operator, only its two degenerate ground states
\begin{equation}
\begin{array}{llll}
|\phi_{0}\rangle= \frac{1}{\sqrt{2+q^2}}(|110\rangle+q
|101\rangle+|011\rangle),
\\
|\phi_{0}'\rangle= \frac{1}{\sqrt{2+q^2}}(|100\rangle+q
|010\rangle+|001\rangle)
\end{array}
\end{equation}
are involved, with $|0\rangle$ and  $|1\rangle$ being the eigenstates of
$\sigma^{z}$ and $q=-\frac{1}{2}(\Delta+\sqrt{\Delta^2+8})$.

Projecting $H$ into this lowest energy subspace, the effective Hamiltonian
up to the first-order correction is obtained to be
$$
H_{eff}=P_{0}(H_{B}+H_{BB})P_{0},
$$
Here $P_{0}$ is a projection operator $P_{0}=\prod^{N/3}_{i=1} P^{L}_{0}$, with
$ P^{L}_{0}=|\Uparrow \rangle \langle \phi_{0} |+|\Downarrow \rangle \langle \phi_{0}' |$
being the $L$-block operator, and $|\Uparrow \rangle$ and $|\Downarrow \rangle$ denoting the new spin-$\frac{1}{2}$ states. The renormalized effective Hamiltonian preserves the same form as Eq.~(\ref{ham1})
\begin{equation}
\begin{array}{llll}
H_{eff}=\frac{1}{4}J'
\sum_{i}^{\frac{N}{3}}(\sigma_{i}^{x}\sigma_{i+1}^{x}+
\sigma_{i}^{y}\sigma_{i+1}^{y}+ \Delta'
\sigma_{i}^{z}\sigma_{i+1}^{z}),
\end{array}
\end{equation}
with the scaled coupling parameters \cite{Kargarian08a}
\begin{equation}
\begin{array}{llll}
J'=J (\frac{2q}{q^2+2})^2,~~~
\Delta'= \frac{\Delta}{4} q^2.
\end{array} \label{iterrel}
\end{equation}
For the RG method, the fixed points of RG flows play a key role. By solving the
equation $\Delta'=\Delta$, we find two fixed points $\Delta=0$ and $1$.
Only $\Delta=1$ is the nontrivial fixed point. Although this
result is different from the critical line $0\leq \Delta \leq 1$
obtained by Bethe Ansatz, if appropriate boundary
terms are adopted, the renormalized method can
correctly predict the critical line \cite{mamdg}.

Fig.~\ref{evoldelta} shows the variation of the renormalized $\Delta'$ with respect to
$\Delta$. It's obvious that $\Delta'$ have different characters
at the two sides of critical point $\Delta=1$.
In the spin-liquid phase $0\le \Delta < 1$,
$\Delta'$ decreases slowly with the increase of the renormalized time $n_{R}$
and approaches to two fixed points $\Delta$=0 and 1. While in the
N\'{e}el phase $\Delta > 1$, it almost exponentially increases.

\begin{figure}[!tb]
\centering
\includegraphics[width=2.2in]{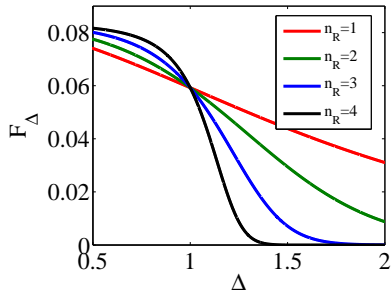}
\hspace{2.2cm}\caption{The variation of the QFI $F_{\Delta}$ with
respect to $\Delta$ in terms of the renormalized time. Here
$n_{R}$ denotes the renormalized time.}
\label{spinchainFSA1}
\end{figure}

\begin{figure}[!tb]
\centering
\includegraphics[width=2.1in]{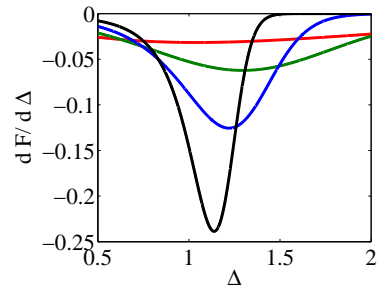}
\hspace{2.2cm}\caption{The variation of $\frac{ d F_{\Delta} }{d
\Delta}$ with respect to $\Delta$ in terms of the renormalized
time.
Each line is one-to-one correspondence to that in Fig.~\ref{spinchainFSA1}. }
\label{spinchainFSA2}
\end{figure}

\section{QFI for renormalized spin-$\frac{1}{2}$ XXZ chain}
\subsection{Quantum Fisher information}
For a given quantum state $\rho_{\varphi}$, its
QFI with respect to parameter $\varphi$ can be defined as \cite{Pairs09}
\begin{equation}
\begin{array}{llll}
F_{\varphi}= \mathrm{Tr} ( \rho_{\varphi} L_{\varphi}^2 ),
\end{array} \label{deffisher}
\end{equation}
where $\varphi$ is the parameter to be estimated,
$L_{\varphi}$ is the symmetry logarithmic derivative, determined by
\begin{equation}
\begin{array}{llll}
\frac{ \partial \rho_{\varphi} }{ \partial \varphi }=\frac{1}{2} (\rho_{\varphi}
L_{\varphi}+ L_{\varphi} \rho_{\varphi} ).
\end{array}
\end{equation}
According to quantum Cram\'{e}r-Rao inequality, the parameter precision has a lower
bound limit
\begin{equation}
\begin{array}{llll}
\Delta \hat{\varphi} \geq \frac{1}{ \sqrt{M F_{\varphi}}},
\end{array}
\end{equation}
where $\hat{\varphi}$ is an unbiased estimator $\mathrm{Tr} (\hat{\varphi} \rho_{\varphi})=\varphi$,
$M$ is the number of trials.

Making use of the spectrum decomposition $\rho_{\varphi}= \sum_{k} \lambda_{k}|k\rangle \langle k|$,
its QFI can be divided into two parts
\begin{equation}
F_{\varphi}= F_{C}+F_{Q}
\label{FisherA}
\end{equation}
with
\begin{equation}
\begin{array}{llll}
F_{C}= \sum_{k} \frac{(\partial_{\varphi }\lambda_{k})^2}{\lambda_{k}},
~~~
F_{Q}=2\sum_{k, k'} \frac{(\lambda_{k}-\lambda_{k'})^2}{\lambda_{k}+\lambda_{k'}} |\langle k | \partial_{\varphi}k' \rangle|^2.
\end{array}  \label{FisherA2}
\end{equation}
Here $\lambda_{k}>0$ and $\lambda_{k}+\lambda_{k'}>0$.
$F_{C}$ is just the classical Fisher information
if one considers $\lambda_{k}$ playing the similar role as the probability.
$F_{Q}$ can be regarded as the pure quantum contribution, since
the factor $|\langle k | \partial_{\varphi}k' \rangle|$ illustrates the quantum coherence
between the eigenvectors of $\rho_{\varphi}$.

\begin{figure}[!tb]
\centering
\includegraphics[width=2.1in]{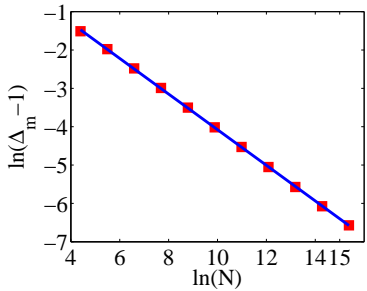}
\hspace{2.2cm}\caption{The scaling behavior of $\Delta_{m}$ in terms of the
system size $N$.
The square is the numerical result and the straight line is the fitting one.}
\label{spinchainFSA3}
\end{figure}

\subsection{QFI of full $N$ sites}
In the section II, the RG method
is applied to the spin-$\frac{1}{2}$ XXZ chain. This method can
effectively rescale the size of large system ($N=3^{n_{R}+1}$)
to a three sites with the renormalized parameters after the $n_{R}$-th RG
iteration. In order to adopt the QFI to characterize the BKT transition
of the XXZ model, one of its renormalized ground states of the full $N$ sites
is considered. Correspondingly, the density matrix is
\begin{equation}
\begin{array}{llll}
\rho_{123}=|\phi_{0}\rangle\langle \phi_{0}|.
\end{array}\label{initstate}
\end{equation}
We have checked that choosing $|\phi_{0}'\rangle$ yields the same results.
By the direct diagonalization,
the eigenvectors and the corresponding eigenvalues of $\rho_{123}$ are obtained
(They are omitted here). As the eigenvalues of $\rho_{123}$ are independent of $\Delta$,
according to Eq.~(\ref{FisherA2}), the QFI of $\rho_{123}$ only has quantum part $F_{Q}$,
without classical contribution $F_{C}$. Straightforward calculation gives its QFI with respect to $\Delta$
\begin{equation}
\begin{array}{llll}
F_{\Delta}=\frac{4 q^2}{(8 + \Delta^2)(3 - \Delta q)(4-\Delta q)}.
\end{array}\label{FisherresutA}
\end{equation}

Firstly, we consider the asymptotic limits of $F_{\Delta}$
on the two sides of the critical point $\Delta=1$.
In the limit $\Delta\rightarrow 0$, $F_{\Delta}=\frac{1}{12}$.
It is a constant independent of $\Delta$ or $n_{R}$.
On the other hand, in the limit $\Delta \rightarrow \infty$,
\begin{equation}
\begin{array}{llll}
F_{\Delta}=\frac{4}{\Delta^4}.
\end{array}
\label{FisherresutA2}
\end{equation}
With the increase of the renormalized time $n_{R}$,
$F_{\Delta}$ converges to zero very rapidly with the
increase of $\Delta$ (shown in Figs.~\ref{evoldelta}).

Next, the numerical simulations are used to study the other behaviors of $F_{\Delta}$.
In Fig.~\ref{spinchainFSA1}, we plot the variation of $F_{\Delta}$ with respect to $\Delta$.
Each curve crosses at the critical value $\Delta=1$.
Corresponding to the different phases, QFI has two saturated values. In the
the spin-liquid phase $0\leq \Delta \leq 1$, $F_{\Delta} \simeq 0.08 \simeq \frac{1}{12}$,
while in the N$\acute{e}$el phase $\Delta > 1$, $F_{\Delta}\rightarrow 0$.

The physics of these results can be understood as following.
The competition between the first two terms and the third term in Eq.~\ref{ham1}
determines the main property of the XXZ model near the critical point.
In the spin-liquid phase, the first two terms of Eq.~\ref{ham1} dominate. Then, a small change of
$\Delta$ in this phase can changes the ground state of system considerably by
affecting the third term of Eq.~\ref{ham1}. So QFI in this phase has relatively large value.
On the other hand, in the N$\acute{e}$el phase,
the roles between the first two terms and the third term reverse: the last term has the main
contribution. Therefore, the ground state is not so sensitivity as that in the spin-liquid phase,
and the value of QFI is small.

\begin{figure}[!tb]
\centering
\includegraphics[width=2.1in]{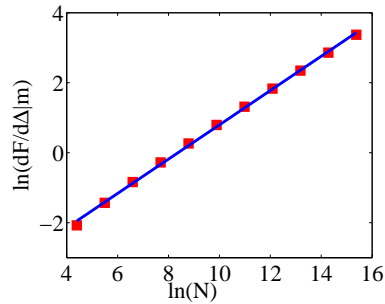}
\hspace{2.2cm}\caption{$\ln(\frac{ d F_{max} }{d \Delta})|_{m}$ with respect
to $\ln(N)$. The squares are the numerical results and the straight line is the fitting one.}
\label{spinchainFSA4}
\end{figure}

The non-analytical property of QFI, which associates with the finite-size scaling laws,
is studied by the first derivative of $F_{\Delta}$ with respect to $\Delta$, as shown in
Fig.~\ref{spinchainFSA2}. This figure clearly displays a discontinuous at $\Delta=1$.
With the increase of system size, the minimum value of $\frac{ d F_{\Delta} }{d \Delta}$
decreases, and the position $\Delta_{m}$ corresponding to the minimum value
of $\frac{ d F_{\Delta} }{d \Delta}$, which can be considered as the
pseudo-critical point, approaches to the true critical $\Delta_{c}=1$.
The finite-scaling laws of the QFI are shown in Figs.~\ref{spinchainFSA3} and \ref{spinchainFSA4},
which are obtained by fitting the relationships between
$(\Delta_{m}-1)$ and $\frac{ d F_{\Delta} }{d \Delta}|_{m}$
verse the size of the system $N(=3^{n_{R}+1})$. The numerical results are
\begin{equation}
\begin{array}{llll}
\frac{ d F }{d \Delta}|_{m} \propto N^{\beta},~~~
(\Delta_{m}-1) \propto N^{-\beta},
\end{array}
\end{equation}
with $\beta=0.47$.
It equals the critical exponent obtained by
the renormalized concurrence \cite{Kargarian08a} and discord \cite{zfh12}.

\subsection{QFI of a reduced $\frac{N}{3}$ sites}
In the upper subsection, we study the groud state KT QPT of the whole system.
It is well-known that approaching the critical point,
the whole system is strongly corrected and highly self-similar.
Therefore, it is expected that the reduced state of the whole system
should also signify the critical property of the system. As the $\frac{2N}{3}$ sites can be
effectively reduce to a two-qubit subsystem and it has been investigated by concurrence and
discord, in this subsection, we only focus on a reduced $\frac{N}{3}$ sites.

Further tracing over the sites $2$ and $3$, the resulting reduced density matrix of site $1$
is to be
\begin{equation}
\rho_{1}=
\frac{1}{2+q^2} \left(                 
\begin{array}{cc}   
1+q^2 & 0\\  
0 &   1 \\  
\end{array}
\right).               
\end{equation}

It's easy to see that the eigenvectors of $\rho_{1}$ are independent of $\Delta$. So
the QFI of $\rho_{1}$ with respect to $\Delta$ only has the classical part, given as
\begin{equation}
\begin{array}{llll}
F_{\Delta, 1}=\frac{8q}{(8+\Delta^2)^2 (\Delta+3q)}.
\end{array}
\end{equation}
In the limit $\Delta\rightarrow 0$, $F_{\Delta, 1}=\frac{1}{24}$.
On the other hand, in the limit $\Delta \rightarrow \infty$,
$F_{\Delta, 1}=\frac{4}{\Delta^{4}}$, which equals to the Eq.~(\ref{FisherresutA2}).
We also study the variation of $F_{\Delta, 1}$ with respect to $\Delta$
by the numerical simulations (the numerical results are omitted here).
Although its mathematical form is distinct from Eq.~(\ref{FisherresutA}),
it demonstrates highly similar behaviors as that of Fig.~\ref{spinchainFSA1}.
One clue for this similarity is that they display almost the same
behaviors in the two limits $\Delta\rightarrow 0$ and $\infty$.

In order to improve the precision of parameter estimation, i.e., QFI,
an entangled state is usually adopted.
For our $\frac{N}{3}$ sites subsystem reduced from a pure ground state,
the entanglement can be quantified by entropy.
The von Neumann entropy of $\rho_{1}$ is given as
\begin{equation}
\begin{array}{llll}
S_{1}=-\mathrm{Tr}(\rho_{1} \log \rho_{1}) =-\frac{1}{2+q^2} \log (\frac{1}{2+q^2})-\frac{1+q^2}{2+q^2} \log (\frac{1+q^2}{2+q^2}).
\end{array}
\end{equation}

\begin{figure}[!tb]
\centering
\includegraphics[width=2.0in]{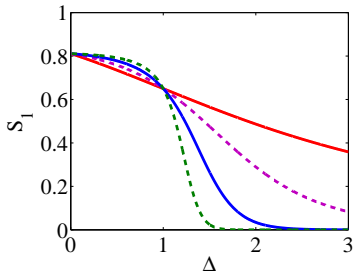}
\hspace{2.2cm}\caption{Entropy $S_{1}$ with respect
to $\Delta$.
From up to bottom, the renormalized times vary from 1 to 4 correspondingly.
It resembles to Fig.~\ref{spinchainFSA1}.}
\label{entropy}
\end{figure}

In Fig.~\ref{entropy}, we show the behaviors of $S_{1}$ as a function of $\Delta$.
It's obvious that it is similar to that of Fig.~\ref{spinchainFSA1}. Every curve
crosses at $\Delta=1$, and the values of $S_{1}$ in the spin-liquid phase $0\leq \Delta \leq 1$
is larger than that of in the
N$\acute{e}$el phase $\Delta > 1$. In a short, a state with a larger entanglement
may enhance the precision of parameter estimation. The similar conclusion has been
obtained for a Bose-Einstein condensates in a symmetric double well \cite{wzhong12a}.

\subsection{Discussion}
A cornerstone of QPT is \textit{universality} in
which the critical behavior depends only on the dimension
of the system and the symmetry of the order parameter \cite{aalt06}.
This universality implies that the critical exponent
is independent of the adopted physical quantity to probe QPT.
By numerical simulation, we also find that the critical exponents in
the finite-size scaling behaviors of $F_{\Delta, 1}$ and $S_{1}$
are also $\beta=0.47$.
It also equals the critical exponent obtained by the method of the renormalized concurrence \cite{Kargarian08a}
and discord \cite{zfh12}. These results demonstrate that the critical exponent
$\beta=0.47$ of XXZ model is universal.
Compared to the renormalized concurrence \cite{Kargarian08a} or discord \cite{zfh12},
the advantage of the QFI is that one can choose one, two or three effectively sites
to characterize the BKT QPT, in contrast to the two-qubit case for
the renormalized concurrence or discord.

However, there is a disadvantage for our RG method. Actually, it is well known
that QFI has two bounds \cite{wiseman10} (a) $F \approx N$ corresponding to
shot-noise limit and (b) $F \approx N^{2}$ to Heisenberg limit.
The numerical result of QFI for our renormalized of spin chain, taking $n_{R}=4$ as an example,
equals about $0.08$, as shown in
Fig.~\ref{spinchainFSA1}. It is much smaller than the shot-noise limit
$F \approx N=3^{5}$. The reason for this distinct difference should be that our
RG scheme reduces the system size just by parameter iteration for a fixed site block.

\section{Conclusion}
To summarize, we have analyzed Berezinskii-Kosterlitz-Thouless quantum phase transition of
spin-$\frac{1}{2}$ XXZ chain using the quantum
renormalized group method. Quantum Fisher information, which is
related to Cram\'{e}r-Rao inequality in quantum estimation
theory, is adopted to detect the critical point.
The quantum Fisher information of the whole and partial system
can equivalently display the quantum phase
transition of this model.
The finite size scaling laws near the critical point is also
investigated, and a universality the critical exponent $\beta=0.47$
is verified. In a word,
quantum Fisher information can be used to
characterize Berezinskii-Kosterlitz-Thouless quantum phase transition.

\textbf{Acknowledgements}
We thanks the helpful discussion with Prof. X. G. Wang and Drs. S. W. Li and
L. Ge. This work is partially supported by the National Natural Science
Foundation of China (Grant Nos.~11065005 and 11365006).



\end{document}